\documentclass[11pt, superscriptaddress, longbibliography,
  nofootinbib]{revtex4-1}

\usepackage{graphicx} 
\usepackage{xcolor,amsmath} 
\usepackage[colorlinks=true, urlcolor=violet,
  linkcolor=blue, citecolor=red, hyperindex=true,
  linktocpage=true]{hyperref}

\begin{document} 

\title{Shock waves, rarefaction waves and non-equilibrium steady
  states \\ in quantum critical systems}

\author{Andrew Lucas} \email{lucas@fas.harvard.edu}
\affiliation{Department of Physics, Harvard University, Cambridge, MA
  02138 USA}

\author{Koenraad Schalm} \email{kschalm@lorentz.leidenuniv.nl}
\affiliation{Instituut-Lorentz for Theoretical Physics, Leiden
  University, Niels Bohrweg 2, Leiden 2333 CA, The Netherlands}

\author{Benjamin Doyon} \email{benjamin.doyon@kcl.ac.uk}
\affiliation{Department of Mathematics, King's College London, Strand,
  London WC2R 2LS, United Kingdom}

\author{M. J. Bhaseen} \email{joe.bhaseen@kcl.ac.uk}
\affiliation{Department of Physics, King's College London, Strand,
  London WC2R 2LS, United Kingdom}

\begin{abstract}
We re-examine the emergence of a universal non-equilibrium steady
state following a local quench between quantum critical heat baths in
spatial dimensions greater than one. We show that energy transport
proceeds by the formation of an instantaneous shock wave and a
broadening rarefaction wave on either side of the interface, and not
by two shock waves as previously proposed. For small temperature
differences the universal steady state energy currents of the
two-shock and rarefaction-shock solutions coincide.  Over a broad
range of parameters, the difference in the energy flow across the
interface between these two solutions is at the level of two
percent. The properties of the energy flow remain fully universal and
independent of the microscopic theory. We briefly discuss the width of
the shock wave in a viscous fluid, the effects of momentum relaxation,
and the generalization to charged fluids.
\end{abstract}

\maketitle

\section{Introduction}

In recent years there has been intense experimental and theoretical
activity exploring the behavior of non-equilibrium quantum systems
\cite{Polkovnikov:RMP}. Stimulated by experiment on low-dimensional
cold atomic gases \cite{Kinoshita:Newton}, theoretical work has
focused on the dynamics of integrable models and their novel
thermalization properties.  An important finding is that integrable
models are typically described by a Generalized Gibbs Ensemble (GGE)
\cite{Rigol:GGE,Rigol:Therm,Rigol:Breakdown} due to the presence of an
infinite number of conservation laws. However, there are very few
theoretical results in non-integrable settings and in higher
dimensions. Recent experiments on cold atomic gases \cite{schafer},
Fermi liquids \cite{molenkamp, geim, mackenzie}, and charge neutral
graphene \cite{kim1, kim2}, probe the dynamics of quantum systems in
more than one dimension. It is timely to establish universal phenomena
for such higher dimensional systems.

In recent work we investigated non-equilibrium energy transport
between quantum critical heat baths in arbitrary dimensions
\cite{Bhaseen:Energy}, generalizing the results of \cite{Doyon:Heat}
for one spatial dimension.  We showed that a non-equilibrium steady
state (NESS) emerges between the heat baths and that it is equivalent
to a Lorentz boosted thermal state. The latter captures both the
average energy current and its fluctuations. In particular, the energy
current and its entire fluctuation spectrum is universally determined
in terms of the effective ``central charge'' (the analogue of the
Stefan-Boltzmann constant) of the quantum critical heat baths and
their temperatures. A key observation is that the steady state is
formed by propagating wavefronts emanating from the contact
region. For small temperature differences these wavefronts are
ordinary sound waves, but for large temperature differences their
dynamics is non-linear. The properties of the NESS are constrained by
the equation of state of the heat baths and the conservation of energy
and momentum across the wavefronts. This hydrodynamic approach based
on macroscopic conservation laws thus provides a valuable handle on
non-equilibrium transport in arbitrary dimensions, establishing
bridges between different fields of research. The emergence of a NESS
bounded by two planar shock waves was also considered in
Refs.~\cite{Chang:Ansatz,Amado:Steady,Pourhasan:NESS}.

In this paper we re-examine this problem of non-equilibrium energy
flow in arbitrary dimensions. We show that the idealized solution in
terms of two infinitely sharp shock waves requires modification in the
light of thermodynamic and numerical considerations.  In spatial
dimensions $d>1$, one of the shocks is actually a smoothly varying and
broadening rarefaction wave, even in the absence of viscosity. The
results in $d=1$ are unaffected due to the light-cone propagation of
the wavefronts, where the effective speed of light and the speed of
sound coincide. In higher dimensions this is not the case and more
complicated solutions may arise. Even in the presence of a broad
rarefaction wave in $d>1$, we always find that a NESS is supported at
the interface between the heat baths. This NESS can once again be
understood as a Lorentz-boosted thermal state. In particular,
numerical and analytical results for the solutions show that the
steady state energy current is again universal. Quantitatively the
effect of a broadening rarefaction wave is small: for the
experimentally relevant dimensions of $d=2,3$, the results for the
energy current across the interface agree to within about $2\%$ of the
idealized sharp shock solution over a broad range of temperatures. In
physically realizable systems, shock broadening will also occur due to
viscous corrections ~\cite{Bhaseen:Energy}. We outline the
non-perturbative effects of this broadening in Section
\ref{Sec:Viscous}. We also provide a brief discussion of momentum
relaxation in Section \ref{Sec:Momentum} and of charged fluids in
Section \ref{Sec:Charge}.  We conclude in Section \ref{Sec:Conc} with
an outlook for future research.

\section{Universal NESS between quantum critical heat baths}
\label{Sec:Setup} 

\begin{figure}[t!]
\begin{center}
\includegraphics[width=0.45\textwidth,clip=true]{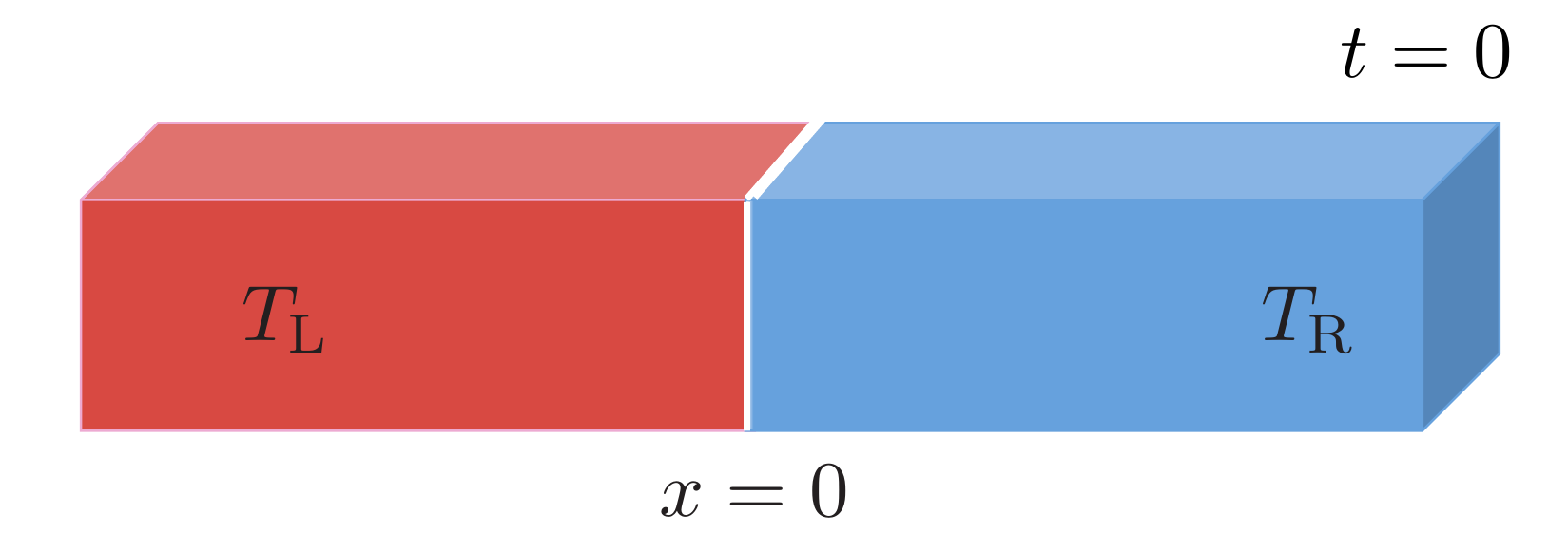}
\end{center}
\caption{The set-up consists of two isolated quantum critical systems
  which are initially prepared at temperatures $T_{\mathrm{L}}$ and
  $T_{\mathrm{R}}$ and are instantaneously connected along a
  hyperplane at time $t=0$. A non-equilibrium steady state (NESS)
  forms at the interface between the heat baths carrying an average
  energy current $J_{\mathrm{E}}$. Within a hydrodynamic approach
  based on macroscopic conservation laws, the character of the NESS is
  determined by the equation of state of the heat baths and
  energy-momentum conservation across the resulting wavefronts. The
  latter may take the form of sharp shock waves or smoothly varying
  rarefaction waves, depending on the spatial dimensionality.}
\label{Fig:Setup}
\end{figure}

The set-up we consider is depicted in Fig.~\ref{Fig:Setup}. Two
infinitely large isolated but identical quantum critical systems are
initially prepared at temperatures $T_{\mathrm{L}}$ and
$T_{\mathrm{R}}$ and are brought into instantaneous contact along a
hyperplane at time $t=0$ \cite{Doyon:Heat,Bhaseen:Energy}.  We
restrict our attention to Lorentz invariant quantum critical points
with an effective speed of light $v_l=1$.  On connecting the two
systems together, a NESS forms at the interface between the heat
baths, carrying a ballistic energy current $J_{\mathrm{E}}= T^{tx}$,
where $T^{\mu\nu}$ is the energy-momentum tensor.\footnote{Here and
  henceforth, we implicitly average over quantum and thermal
  fluctuations when defining the stress tensor.} This ``partitioning''
setup may be regarded as a local quantum quench joining two
independent sub-systems. Equivalently, we may consider applying an
abrupt step temperature profile to an otherwise uniform system.  In
the context of hydrodynamics these initial conditions correspond to
the so-called Riemann problem, to which we will return in Section
\ref{Sec:Higher}.

\bigskip

Let us briefly recall the results in $d=1$. In one spatial dimension a
spatially homogeneous NESS is formed in the vicinity of the interface
\cite{Doyon:Heat,Bernard:Time,Bernard:Noneq}. The steady state carries
a universal average energy current
$J_{\mathrm{E}}=c\pi^2k_{\mathrm{B}}^2(T_{\mathrm{L}}^2-T_{\mathrm{R}}^2)/6h$,
where $c$ is the central charge of the heat baths. This may be
regarded as an application of the Stefan-Boltzmann law to quantum
critical systems, where the internal energy density is proportional to
$T^{d+1}$ \cite{Cardy:Ubiquitous}. The result for $J_{\mathrm{E}}$
extends earlier results for free fermions and bosons
\cite{Pendry:Limits,Sivan:Multi,Butcher:Thermal,Fazio:Anom}, as
confirmed by transport experiments on ballistic channels
\cite{Jezouin:Quantum,Schwab:Measurement,Rego:Quantized}. The
generalization to arbitrary $c$ has been verified by time-dependent
Density Matrix Renormalization Group (DMRG) methods on quantum spin
chains
\cite{Karrasch:Noneqtherm,Karrasch:Finite,Karrasch:Reducing,Huang:Scaling}.
Moreover, the exact generating function of energy current fluctuations
has also been determined \cite{Doyon:Heat,Bernard:Time,Bernard:Noneq}.

\begin{figure}[t!]
\begin{center}
\includegraphics[width=0.45\textwidth,clip=true]{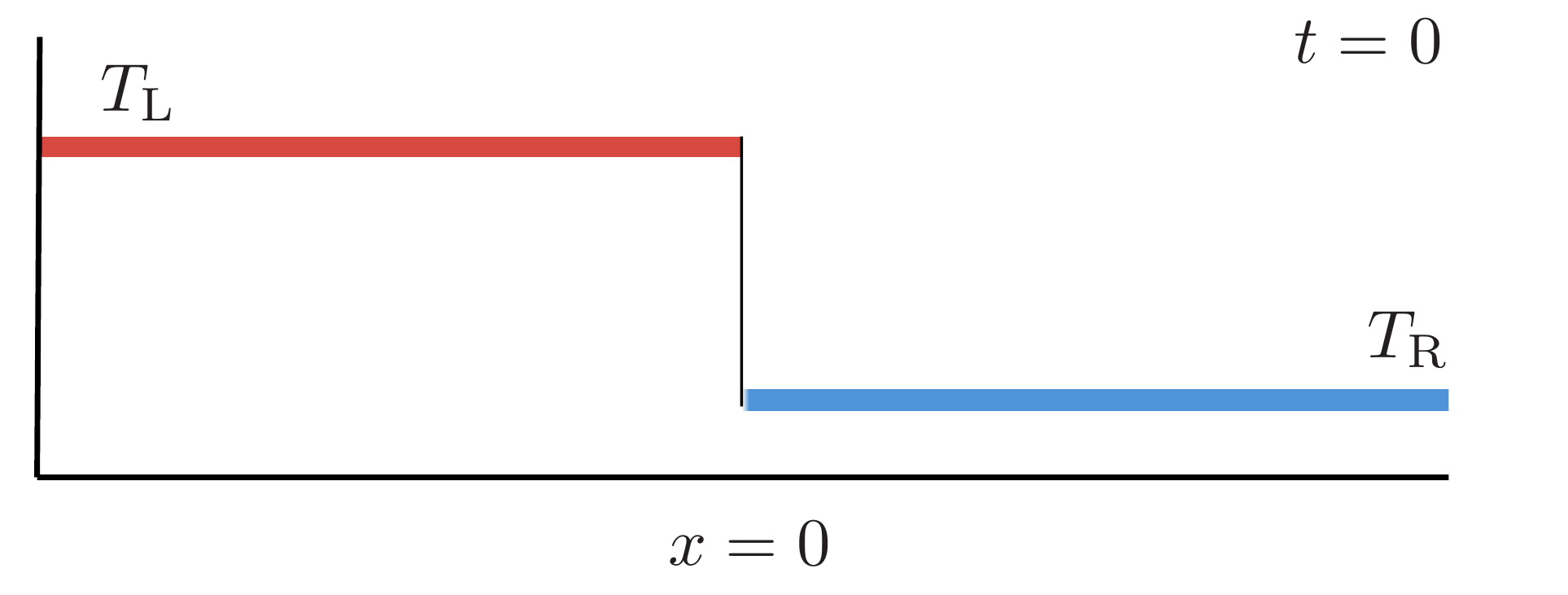}
~
\includegraphics[width=0.45\textwidth]{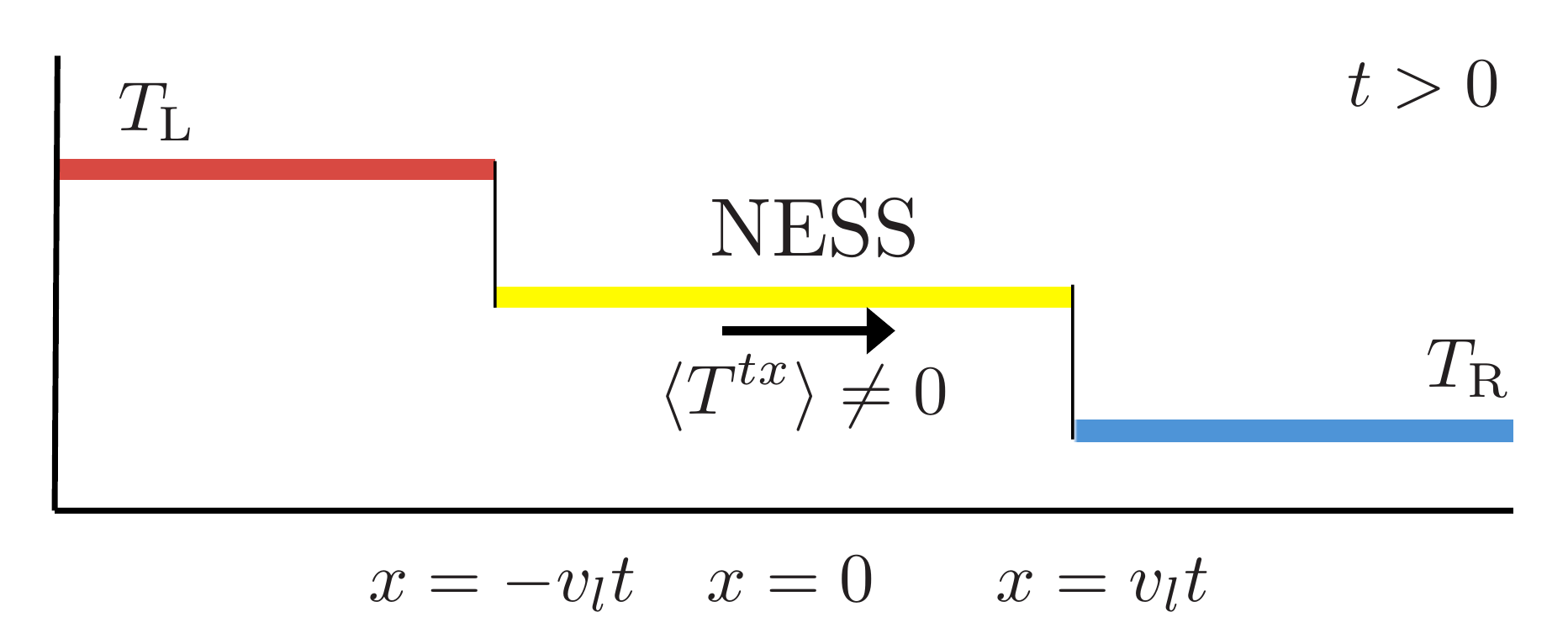}
\end{center}
\caption{In one spatial dimension a spatially homogeneous steady state
  region is formed by outgoing shock waves moving at the effective
  speed of light $v_l$. The energy current $J_{\mathrm{E}}$ and the
  exact spectrum of energy current fluctuations are completely
  described a Lorentz boosted thermal state with temperature
  $T=\sqrt{T_{\mathrm{L}}T_{\mathrm{R}}}$.}
\label{Fig:1D}
\end{figure}

In Ref.~\cite{Bhaseen:Energy} we discussed this non-equilibrium energy
transport problem from a rather different vantage point. By combining
insights from gauge-gravity duality and the dynamics of energy and
momentum conservation, we showed that the 1+1 dimensional NESS is
completely equivalent to a Lorentz boosted thermal state: by
``running'' past a thermal state at temperature
$T=\sqrt{T_{\mathrm{L}}T_{\mathrm{R}}}$ it is possible to reproduce
both the average energy flow {\em and} the full spectrum of energy
current fluctuations in the NESS. Moreover, it is possible to extract
the time-dependence from the solution of the macroscopic conservation
laws $\partial_\mu T^{\mu\nu}=0$ in $1+1$ dimensions. The spatially
homogeneous region is formed by outgoing ``shock waves'' which emanate
from the point of contact at the effective speed of light; see
Fig.~\ref{Fig:1D}.
In particular, the form of the steady state solution is uniquely
determined by energy-momentum conservation across the shock fronts.
This macroscopic conservation law approach is readily generalized to
other equations of state for the energy baths. This has been recently
demonstrated for perturbed 1+1 dimensional CFTs
\cite{Bernard:Hydro}. The use of conservation laws across large
transition regions has also led to a thermodynamic description for the
total, integrated current in one-dimensional systems
\cite{Vasseur:Expansion}. This growing body of work opens the door to
wider applications of hydrodynamic techniques in low-dimensional
quantum systems; for earlier work in this direction see for example
Ref.~\cite{Abanov:Hydro}.

\subsection{The NESS in $d>1$}
\label{Sec:Higher}

\begin{figure}
\begin{center}
\includegraphics[width=0.45\textwidth,clip=true]{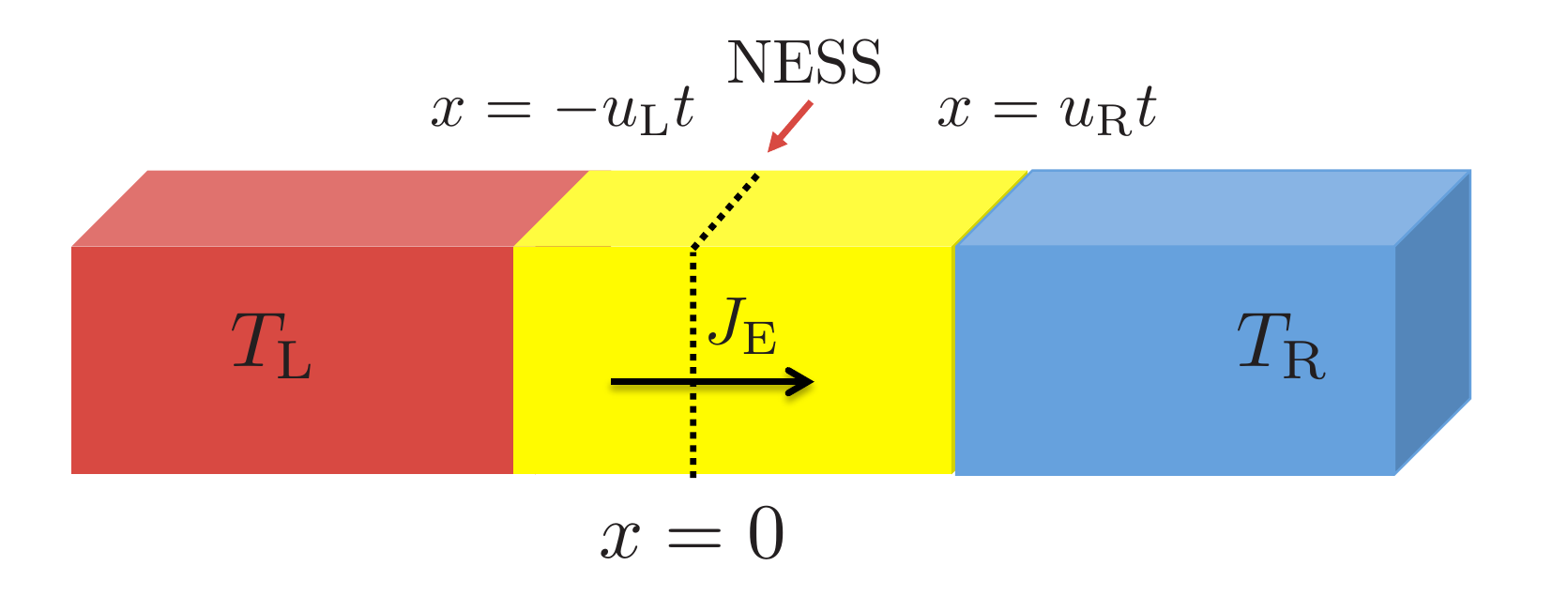}
~ \raisebox{-.005in}{
  \includegraphics[width=0.45\textwidth,clip=true]{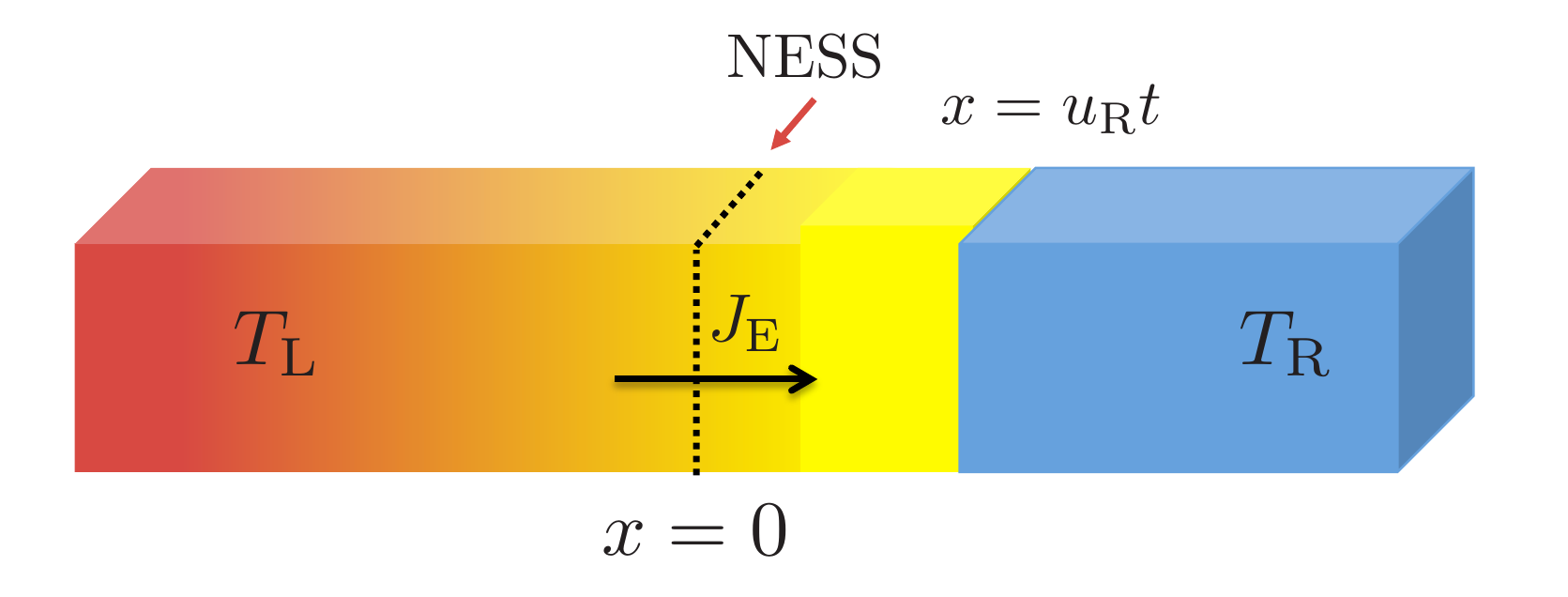}}
\end{center}
\caption{(a) Idealized solutions to conformal hydrodynamics in $d>1$
  consisting of two planar shock waves emanating from the contact
  region. Thermodynamic consistency requires that the left-moving
  shock wave is replaced by a smooth rarefaction wave, even in the
  absence of viscosity. (b) Modified solution consisting of a
  left-moving rarefaction wave and a right-moving shock wave. The
  difference between the average energy current $J_{\rm E}$ across the
  interface in the two cases is about two percent.  Note that a
  spatially homogeneous region also occurs to the right of the
  rarefaction wave as indicated by the solid yellow shading; see
  Fig.~\ref{fig4}.}
\label{Fig:Rarefaction}
\end{figure}

In Ref.~\cite{Bhaseen:Energy} we argued that the above results could
be generalized to higher dimensions by invoking the techniques of
relativistic hydrodynamics. In particular, we showed that the
numerical solution of conformal hydrodynamics leads to a non-trivial
NESS in $d=2$, that is robust to a variety of perturbations.
Moreover, we showed that both the average energy current $J_{\rm E}$
and the shock speeds $u_{\rm L,R}$, were in very good quantitative
agreement with analytical solutions based on idealized two-shock
solutions; see Fig~\ref{Fig:Rarefaction}(a). In this work we re-visit
this two-shock ansatz, which we stressed in Ref.~\cite{Bhaseen:Energy}
is not a unique solution, and show that it is necessary to include
rarefaction waves based on thermodynamic arguments; see
Fig~\ref{Fig:Rarefaction}(b).  We show that this leads to even better
agreement with our numerical simulations. Importantly, the solution
still contains a NESS, and the properties of this NESS can be
determined analytically, though there is a change in the exact results
compared to the idealized two-shock solution. The solution is still
universal, and is solely determined by $T_{\mathrm{L,R}}$ and the
analogue of the central charge of the quantum critical theories.

\subsection{Hydrodynamic Limit}
\label{Sec:Hydro}

As for any interacting theory, with strongly coupled conformal field
theories (CFTs) describing the quantum critical heat baths in
Fig.~\ref{Fig:Setup}, the late-time behavior following the local
quench is expected to be captured by relativistic hydrodynamics.  In
particular, for a strongly coupled fluid at temperature $T$, we expect
that hydrodynamics provides a good description of the non-equilibrium
dynamics of conserved quantities on time scales long compared to
$1/T$.  As the relevant time scale $t\rightarrow \infty$, higher
derivative corrections to the hydrodynamic equations can be neglected
\cite{Bhaseen:Energy}.\footnote{This is because one can rescale
  $x\rightarrow \lambda x$ and $t\rightarrow \lambda t$ with
  $\lambda\rightarrow \infty$; in this limit the viscosity $\eta
  \rightarrow \eta/\lambda$.  If $\eta \rightarrow 0$,
  (\ref{eq:conslaws}) are invariant under this rescaling.  This
  implies that there cannot be any intrinsic timescales to the
  solution to our problem (up to those set by viscous and other higher
  derivative corrections).  We will discuss one minor effect due to
  viscosity in Section \ref{Sec:Viscous}.} Thus the relevant
hydrodynamic equations are the conservation of energy and momentum:
\begin{equation}
\partial_\mu T^{\mu\nu} = 0.  \label{eq:conslaws}
\end{equation}
For a quantum critical state in local equilibrium, we know that
\begin{equation}
T^{\mu\nu} = C T^{d+1} \left[(d+1)u^\mu u^\nu +
\eta^{\mu\nu}\right].  \label{eq:tmunu}
\end{equation}
Here $\eta^{\mu\nu}={\rm diag}(-1,1,\dots, 1)$ is the Minkowski
space-time metric and $u^\mu$ is the local fluid velocity.  This
formula is valid both in the asymptotic baths, and in the emergent
NESS.  The only non-universal part of $T^{\mu\nu}$ is the constant
$C$, which effectively counts the number of degrees of freedom in the
CFT; it can be considered as a generalization of the central charge of
$d=1$ dimensional theories. Bringing two hydrodynamical systems with
$T=T_{\mathrm{L}}$ for $x<0$ and $T=T_{\mathrm{R}}$ for $x>0$ into
contact along a local interface is known as the Riemann problem in
fluid dynamics. We will consider the solutions to this problem below.

\subsection{Two-Shock Solution}
\label{Sec:Twoshock}

The solutions of perfect conformal hydrodynamics are not unique for
$d>1$, in contrast to $d=1$.  Hence, to find a proper solution to the
Riemann problem requires additional physical input. Guided by the
exact shock wave solutions found in $d=1$, we suggested that a NESS
would arise between planar shock waves in $d>1$. Using this ansatz we
argued previously \cite{Bhaseen:Energy} that the non-equilibrium
steady state was equivalent to a Lorentz boosted thermal state at
temperature \begin{equation} T_{\mathrm
    s}=\sqrt{T_{\mathrm{L}}T_{\mathrm{R}}}.
\end{equation}
The corresponding energy flow was given by \cite{Bhaseen:Energy}
\begin{equation}
J_{\mathrm{E}} \equiv T^{tx}_{\mathrm{s}} =
C\frac{T_{\mathrm{L}}^{d+1}-T_{\mathrm{R}}^{d+1}}{v_{\mathrm{R}} +
  (dv_{\mathrm{R}})^{-1}}, \;\;\;\;\; v_{\mathrm{R}} =
\sqrt{\frac{1}{d} \frac{dT_{\mathrm{L}}^{(d+1)/2} +
    T_{\mathrm{R}}^{(d+1)/2}}{dT_{\mathrm{R}}^{(d+1)/2} +
    T_{\mathrm{L}}^{(d+1)/2}}}.
\end{equation}
In particular, this NESS was separated by two asymmetrically moving
shock waves, that can be identified as non-linear sound waves; see
Fig.~\ref{Fig:Rarefaction}(a). Checking this ansatz against numerical
simulations of conformal hydrodynamics we found very good agreement
with our analytical prediction for $J_{\mathrm{E}}$, even far from the
linear response regime.

In spite of this agreement, this solution is problematic for the
following physical reason.  If we truncate the hydrodynamic gradient
expansion at zeroth order in derivatives (perfect hydrodynamics), then
Eq.~(\ref{eq:conslaws}) implies conservation of
entropy \begin{equation} \partial_\mu \left((d+1)C T^d u^\mu\right)
  \equiv \partial_\mu s^\mu = 0,
\end{equation}
on any smooth solution.  However, at an infinitely sharp shock wave
this criterion is generally violated. This is not a problem, as long
as $\partial_\mu s^\mu \ge 0$, which is a local statement of the
second law of thermodynamics.  This is a constraint of hydrodynamics
at all orders in the gradient expansion.  At first order for a
conformal fluid, we have $\partial_\mu s^\mu \sim \eta (\partial
v)^2/T$ (schematically).  The fact that viscosity is \emph{required}
to create entropy at a shock front is a subtlety we will return to in
the next section.  Away from these shock fronts, we will have
$\partial_\mu s^\mu = 0$ in perfect fluid dynamics.

Consider now a shock wave moving at velocity $v_{\mathrm{shock}}$,
with $T_<$ and $v_<$ the fluid temperature and velocity to the left of
the shock, and $T_>$ and $v_>$ the fluid temperature and velocity to
the right of the shock.  Then, integrating over a shock of transverse
area $A$ across a time step $t$, we find that \begin{equation}
  \int\limits_{\text{shock}} \mathrm{d}t \mathrm{d}^d\mathbf{x}\;
  \partial_\mu \left((d+1)C T^d u^\mu\right) = At\times
  (d+1)C\left[\frac{T_>^d (v_> - v_{\mathrm{shock}}) }{\sqrt{1-v_>^2}}
    - \frac{T_<^d (v_< - v_{\mathrm{shock}})
    }{\sqrt{1-v_<^2}}\right].  \label{shockentropy}
\end{equation}
By the argument above, on a physical solution, the right hand side
must be positive. However, one can show that for the shock moving into
the region of higher temperature in this two-shock solution (the
left-moving one), the entropy production given by
Eq.~(\ref{shockentropy}) is negative.  We now describe the
modification of this shock wave so that there is no local entropy
loss.

\subsection{Rarefaction Waves}
\label{Sec:Rarefaction}

Because only the left-moving shock violates the second law of
thermodynamics, we will look for a different solution to the Riemann
problem where the left-moving shock is replaced with a left-moving
rarefaction wave; see Fig.~\ref{Fig:Rarefaction}(b). This is a
solution that is continuous, but whose first derivatives are
discontinuous \cite{taub, thompson, lax, liu, MartiMueller}, and where $T$ and
$v\equiv u^x/u^t$ are functions of $x/t\equiv \xi$ alone. By
assumption therefore the local configuration is always in local
equilibrium, in contrast to a true shock. Very similar solutions were
presented in \cite{mach}. The non-trivial equations of hydrodynamics
are the $t$ and $x$ components of (\ref{eq:conslaws}), and may be
expressed as ordinary differential equations in
$\xi$: \begin{subequations}\begin{align} \xi
    \frac{\mathrm{d}}{\mathrm{d}\xi} \left(T^{d+1}
    \frac{d+v^2}{1-v^2}\right) &= \frac{\mathrm{d}}{\mathrm{d}\xi}
    \left(T^{d+1} \frac{(d+1)v}{1-v^2}\right) ,
    \\ \xi\frac{\mathrm{d}}{\mathrm{d}\xi} \left(T^{d+1}
    \frac{(d+1)v}{1-v^2}\right) &= \frac{\mathrm{d}}{\mathrm{d}\xi}
    \left(T^{d+1} \frac{1+dv^2}{1-v^2}\right).
\end{align}\end{subequations}
Note that the coefficient $C$ of the local equilibrium configuration
drops out, and the solution for the rarefaction profile is independent
of the value of this parameter.

As in \cite{mach}, this pair of equations can be re-organized into the
form
\begin{equation} \left(\begin{array}{c} 0 \\ 0 \end{array}\right)
  = \mathsf{M}(\xi)\left(\begin{array}{c} \mathrm{d}T/\mathrm{d}\xi
    \\ \mathrm{d}v/\mathrm{d}\xi\end{array}\right),
\end{equation}
and is thus only satisfied when $\det(\mathsf{M}(\xi)) = 0$.  A
straightforward calculation reveals that this occurs
when
\begin{equation}
  \left((d+1)v-\left(d+v^2\right)\xi\right)\left(2v-\left(1+v^2\right)\xi\right)
  = \left(1+v^2-2v\xi\right)\left(1+dv^2-(d+1)v\xi\right).
\end{equation}
After further algebraic manipulations, we find that this occurs when
\begin{equation} \xi = \frac{v\pm c_{\mathrm{s}}}{1\pm
  c_{\mathrm{s}}v}, \label{xifromv}
\end{equation} where \begin{equation}
c_{\mathrm{s}} = \frac{1}{\sqrt{d}}
\end{equation}
is the speed of sound in a scale invariant quantum critical
fluid. Eq.~(\ref{xifromv}) is the relativistic velocity addition law
between $\pm c_{\mathrm{s}}$ and the local fluid velocity in the
$x$-direction, $v$. A straightforward inversion reveals
that \begin{equation} v = \frac{\xi \mp c_{\mathrm{s}}}{1\mp
    c_{\mathrm{s}}\xi}.  \label{xifromv2}
\end{equation}
As this left-moving rarefaction wave should begin ($v=0$) when $\xi <
0$, we conclude that we must take the minus sign in
Eq.~(\ref{xifromv}) and the plus sign in Eq. (\ref{xifromv2}).  Next,
we employ entropy conservation in the rarefaction wave, and
obtain \begin{equation} \xi\frac{\mathrm{d}}{\mathrm{d}\xi} \left(
  \frac{T^d}{\sqrt{1-v^2}}\right) = \frac{\mathrm{d}}{\mathrm{d}\xi}
  \left( \frac{T^d v}{\sqrt{1-v^2}}\right).
\end{equation}
Using the relation between $v$ and $\xi$ in a left-moving rarefaction
wave, we convert this equation into a differential equation for
$\mathrm{d}T/\mathrm{d}v$, which may be solved exactly.  Employing the
boundary conditions $T=T_{\rm L}$ at the left-edge of the rarefaction
wave gives
\begin{equation}
T = T_{\mathrm{L}}\left(\frac{1-v}{1+v}\right)^{1/2\sqrt{d}}.  \label{teq}
\end{equation}

Let us now describe the rarefaction wave.  For $\xi <
-c_{\mathrm{s}}$, the solution is $T=T_{\mathrm{L}}$ and $v=0$.  For
$-c_{\mathrm{s}} < \xi < \xi^*$, the solution is described by the
relations (\ref{teq}) and (\ref{xifromv}).  For $\xi^* < \xi <
u_{\mathrm{R}}$, the solution is given by a homogeneous region at
temperature $T_{\mathrm{h}}$ and boosted by a velocity
$v_{\mathrm{h}}$.  Eq.~(\ref{teq}) implies that these are related
via \begin{equation} T_{\mathrm{h}} = T_{\mathrm{L}}
  \left(\frac{1-v_{\mathrm{h}}}{1+v_{\mathrm{h}}}\right)^{1/2\sqrt{d}}.  \label{tsl}
\end{equation}
At $\xi = u_{\mathrm{R}}$ there is a shock wave, and for $\xi >
u_{\mathrm{R}}$, the temperature is $T_{\mathrm{R}}$ and $v=0$.

\begin{figure}[t]
\centering
\includegraphics[width=6in]{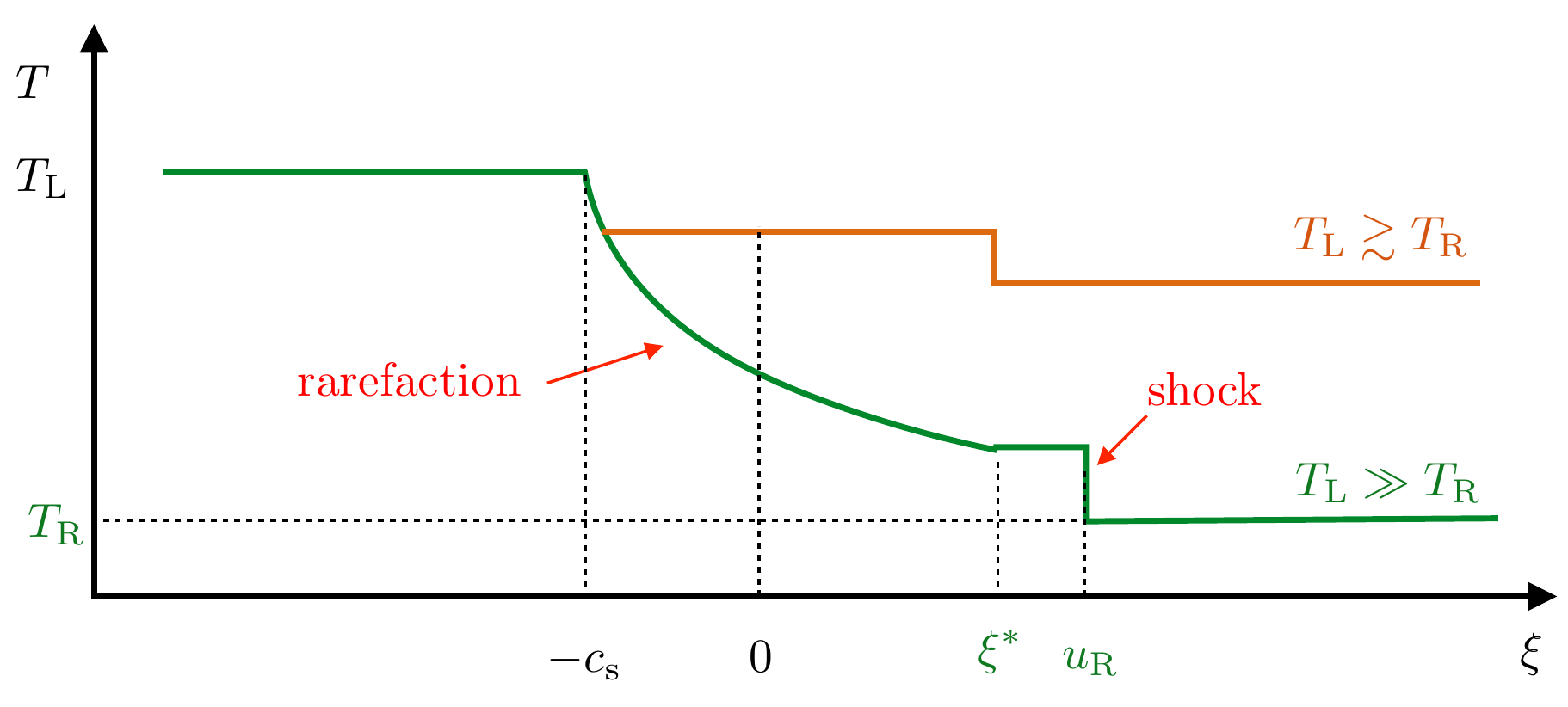}
\caption{A qualitative sketch of the temperature profile $T(\xi)$ in
  the rarefaction-shock solution to the Riemann problem with heat
  baths at temperatures $T_{\mathrm L}$ and $T_{\mathrm R}$. The
  curves correspond to different values of $T_{\mathrm{R}}$ (for fixed
  $T_{\mathrm{L}}$) with $T_{\mathrm L}\gtrsim T_{\mathrm R}$ (orange)
  and $T_{\mathrm L}\gg T_{\mathrm R}$ (green). The profiles coincide
  up until the homogeneous region in between the rarefaction and the
  shock. The location of the latter is dependent on $T_{\mathrm{R}}$.
  For $T_{\mathrm{L}}/T_{\mathrm{R}}<\Gamma$ as given by
  Eq.~(\ref{gammacrit}), a spatially homogeneous profile envelops the
  contact interface (orange), while for
  $T_{\mathrm{L}}/T_{\mathrm{R}}>\Gamma$ the interface resides in the
  rarefaction region (green). A steady state energy current
  $J_{\mathrm E}$ is established at the interface in both cases.}
\label{fig4}
\end{figure}

The complete solution still has the undetermined parameters: $\xi^*$,
$u_{\mathrm{R}}$, $T_{\mathrm{h}}$ and $v_{\mathrm{h}}$.  We can fix
these as follows.  Eq. (\ref{xifromv}) determines $\xi^*$ from
$v_{\mathrm{h}}$.  We then employ the Rankine-Hugoniot conditions
(corresponding to energy and momentum conservation) at the
right-moving shock wave to obtain
\begin{subequations}
\begin{align}
\frac{(d+1)T_{\mathrm{h}}^{d+1}v_{\mathrm{h}}}{1-v_{\mathrm{h}}^2} - u_{\mathrm{R}} T_{\mathrm{h}}^{d+1} \frac{d+v_{\mathrm{h}}^2}{1-v_{\mathrm{h}}^2}  &=-du_{\mathrm{R}}T_{\mathrm{R}}^{d+1}, \\
\frac{1+dv_{\mathrm{h}}^2}{1-v_{\mathrm{h}}^2} T_{\mathrm{h}}^{d+1} - \frac{(d+1)v_{\mathrm{h}}}{1-v_{\mathrm{h}}^2}u_{\mathrm{R}}T_{\mathrm{h}}^{d+1}&= T_{\mathrm{R}}^{d+1}.
\end{align}
\end{subequations}
These equations fix a relation between $T_{\mathrm{h}}$ and
$v_{\mathrm{h}}$:
\begin{equation}
T_{\mathrm{h}} = T_{\mathrm{R}} \left[\frac{2d +
    \left(1+d^2\right)v_{\mathrm{h}}^2 +
    (d+1)v_{\mathrm{h}}\sqrt{4d+(d-1)^2v_{\mathrm{h}}^2}}{2d\left(1-v^2_{\mathrm{h}}\right)}\right]^{1/(d+1)}.
\label{tsr}
\end{equation}
At this point, $u_{\mathrm{R}}$ is also fixed in terms of
$v_{\mathrm{h}}$ and $T_{\mathrm{h}}$.  We now have two formulas,
Eqs. (\ref{tsl}) and (\ref{tsr}) for $T_{\mathrm{h}}$.  There is a
unique value of $v_{\mathrm{h}}$ which satisfies both, and this
completely fixes our solution.  We provide a qualitative sketch of the
final temperature profile $T(\xi)$ for two different temperature
ratios $T_{\mathrm{L}}/T_{\mathrm{R}}$ in Fig.~\ref{fig4}, clarifying
the role of the parameters defined above.  For an observer at a fixed
position $x$, at late times ($t\rightarrow \infty$) $\xi \rightarrow
0$.  As we will see, the energy current is generically non-vanishing,
and this defines a NESS, which is centered at $x=0$.

As may be seen from Fig.~\ref{fig4}, it is possible for the
rarefaction wave to envelop the contact interface at
$x=0$.\footnote{Similar behavior generically happens in free theories
  \cite{collurafree, doyonfree}, in spite of the different physics.
  Free theories do not contain rarefaction waves, but the temporal
  decay towards the NESS away from the contact region is algebraic, as
  in a rarefaction wave.}  Employing Eq.~(\ref{xifromv}) we see that
this occurs when the local speed in the homogeneous region as measured
in the laboratory rest frame exceeds the speed of sound:
\begin{equation}
v_{\mathrm{h}} > c_s = \frac{1}{\sqrt{d}}.
\end{equation}
This occurs at a critical temperature ratio
\begin{align}
\frac{T_{\mathrm{L}}}{T_{\mathrm{R}}} &= \Gamma \equiv  \left(\frac{\sqrt{d}+1}{\sqrt{d}-1}\right)^{1/2\sqrt{d}} \left(\frac{3d + d^{-1} + (d+1)\sqrt{4+(d-1)^2d^{-3/2}}}{2(d-1)}\right)^{1/(d+1)} \notag \\
&\approx \left\lbrace\begin{array}{ll} 3.459 &\ d=2 \\ 2.132 &\ d=3 \end{array}\right..
\label{gammacrit}
\end{align}
When $T_{\mathrm{L}}/T_{\mathrm{R}} < \Gamma$, the rarefaction wave
does not include the origin, and so the NESS is spatially homogeneous
about $x=0$ at finite time $t$.  When $T_{\mathrm{L}}/T_{\mathrm{R}}
>\Gamma$, the NESS is contained in the rarefaction wave, and only
becomes spatially homogeneous asymptotically as $t\rightarrow \infty$;
see Fig.~\ref{fig4}.

It is interesting that the equations derived above for a rarefaction
wave coincide with the exact (two-shock) results in $d=1$
\cite{Doyon:Heat, Bhaseen:Energy}.  However, we stress that there is
no rarefaction wave in $d=1$.

\subsection{Energy Transport at the Interface}

Having established the rarefaction-shock solution, we now examine the
energy current at the interface $x=0$, the location of the emergent
NESS. The energy current at the interface $J_{\mathrm{E}}= T^{tx}$
follows by computing $T(\xi=0)\equiv T_0$ and $v(\xi=0)\equiv v_0$,
and employing Eq.  (\ref{eq:tmunu}): \begin{equation} J_{\mathrm{E}} =
  (d+1)CT_0^{d+1} \frac{v_0}{1-v_0^2}.
\end{equation}
Consider first the limit where $T_{\mathrm{L}}\approx T_{\mathrm{R}}$.
In this regime, the rarefaction wave does not envelop the origin.  We
find $T_0 \approx (T_{\mathrm{L}}+T_{\mathrm{R}})/2$ and (see Appendix
\ref{AppendixA})
\begin{equation}
J_{\mathrm{E}} \approx  \frac{C\sqrt{d}(d+1)}{2} \left(\frac{T_{\mathrm{L}}+T_{\mathrm{R}}}{2}\right)^d (T_{\mathrm{L}}-T_{\mathrm{R}}).  \label{jesmall}
\end{equation}
The rarefaction-shock and two-shock solutions both reproduce this
result, at leading order in $T_{\mathrm{L}}-T_{\mathrm{R}}$.  When
$T_{\mathrm{L}}/T_{\mathrm{R}} > \Gamma$, the rarefaction wave
envelops the origin and we find a universal result \begin{equation}
  J_{\mathrm{E}} = C \frac{(d+1)\sqrt{d}}{d-1}
  \left(\frac{\sqrt{d}-1}{\sqrt{d}+1}\right)^{(d+1)/2\sqrt{d}}
  T_{\mathrm{L}}^{d+1} .  \label{eq:JE0inf}
\end{equation}
Surprisingly, Eq. (\ref{eq:JE0inf}) is independent of
$T_{\mathrm{R}}$. 

\begin{figure}[t]
\centering
\includegraphics[width=6in]{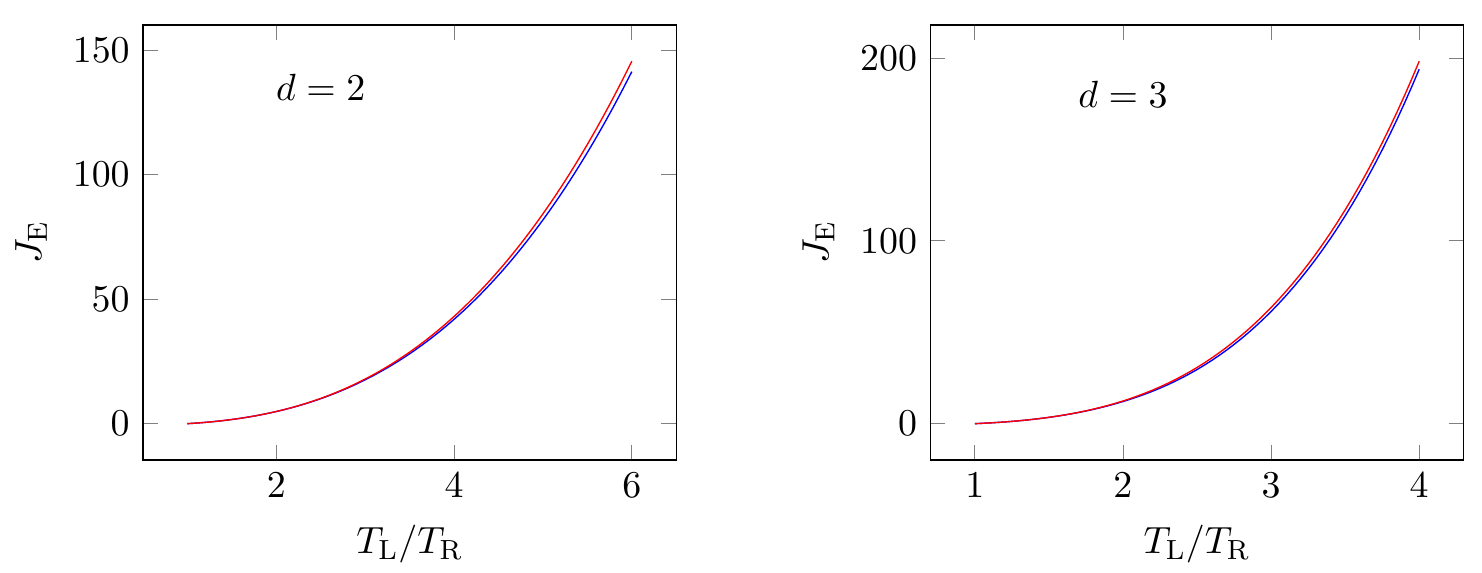}
\caption{A comparison of the rarefaction-shock prediction for
  $J_{\mathrm E}=T^{tx}(x=0)$ (blue), compared to the two-shock
  prediction (red), in $d=2,3$. It is readily seen that results are
  numerically very close to each other. For simplicity, we have
  measured $J_{\mathrm E}$ in units of $CT_{\mathrm{R}}^{d+1}$ in the above plots.}
\label{figexact}
\end{figure}

More generally, we can numerically solve (\ref{tsl}) and (\ref{tsr})
to compute $J_{\mathrm{E}}$ at any
$T_{\mathrm{L},\mathrm{R}}$. Notably, the rarefaction-shock result for
$J_{\mathrm{E}}$ is very close to the one predicted using the
two-shock solution, even as $T_{\mathrm{L}}/T_{\mathrm{R}} \rightarrow
\infty$.  The two predictions are within $2\%$ of each other in the
$T_{\mathrm{L}}/T_{\mathrm{R}}\rightarrow\infty$ limit in both $d=2$
and $d=3$; see Fig.~\ref{figexact}.

\begin{figure}[t]
\centering
\includegraphics[width=6in]{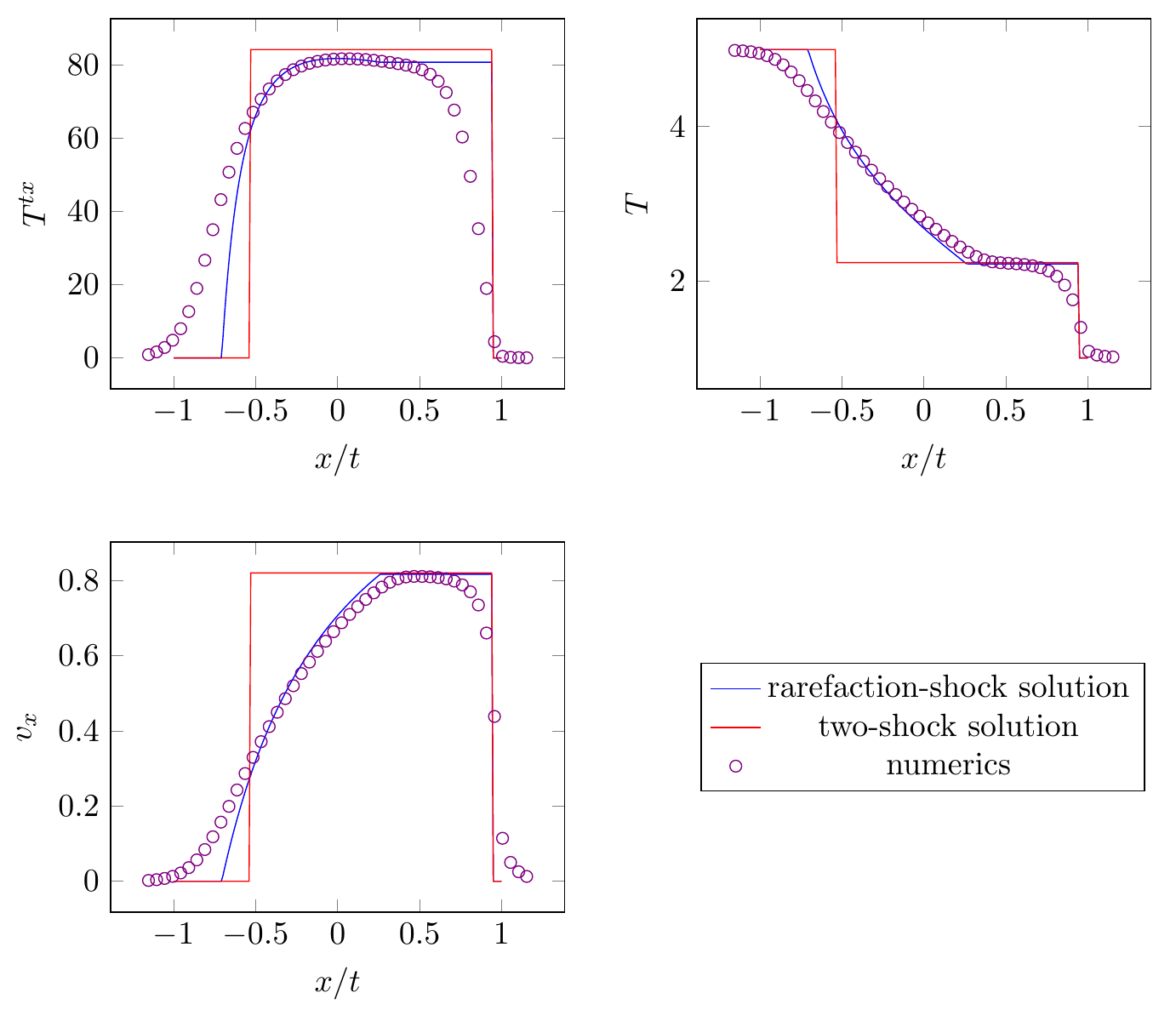}
\caption{A comparison of the hydrodynamic variables $T(x,t)$, $v(x,t)$
  and the resulting energy current, $T^{tx}(t,x)$ in the
  rarefaction-shock solution versus the two-shock solution.  We have
  also included the numerical solution of \cite{Bhaseen:Energy} of the
  Riemann problem (with a smoothed temperature profile) in $d=2$, with
  $T_{\mathrm{L}}=5$ and $T_{\mathrm{R}} = C=1$. The numerical data
  are taken at time $t=1.25$, with initial conditions $T(x,t=0) =
  (T_{\mathrm{L}}+T_{\mathrm{R}})/2 - (T_{\mathrm{L}}-T_{\mathrm{R}})
  \tanh(6.5 \sin(x))$, with periodic boundary conditions at $x=\pm
  \pi$. It is readily seen that the rarefaction-shock solution
  provides a better fit to the data than the two-shock solution.}
\label{fignum}
\end{figure}

The difference between the rarefaction-shock and two-shock solutions
is most transparent in the spatial profile of physical
observables. This is clearly seen in Fig.~\ref{fignum} which compares
the $x$ and $t$ dependence of the rarefaction-shock and two-shock
solutions, to the numerical solution of perfect conformal
hydrodynamics given in Ref.~\cite{Bhaseen:Energy}.  Note that at the
rather extreme pressure ratio of $P_{\mathrm{L}}/P_{\mathrm{R}}>100$
(where $P=C T^{d+1}$ in the fluid rest frame), where the rarefaction
wave envelops the origin, finite size effects are present in our
numerical results.

\section{Viscous Corrections}
\label{Sec:Viscous}

In this section we clarify the qualitative role of dissipative viscous
corrections to the rarefaction-shock dynamics above, and describe the
width of the right-moving shock wave when $T_{\mathrm{L}} >
T_{\mathrm{R}}$.  We focus on $T_{\mathrm{L}}-T_{\mathrm{R}} \ll
T_{\mathrm{R}}$ for simplicity.  Perturbatively, we showed
\cite{Bhaseen:Energy} at intermediate time scales that the shock width
will grow diffusively: $l_{\mathrm{shock}}\sim \sqrt{Dt}$ with
diffusion constant $D\sim \eta T^{-d-1}$ and viscosity $\eta$.  Our
purpose in this section is to expand on this result, and to argue that
perturbation theory fails at late times.

We know from Eq. (\ref{shockentropy}) that the entropy production at
the shock front, per unit transverse area per unit time, is given by
\begin{equation}
\mathcal{S} = (d+1)C\left[\frac{T_{\mathrm{h}}^d (u_{{\mathrm R}} -
    v_{\mathrm{h}})}{\sqrt{1-v_{\mathrm{h}}^2}} - T_{\mathrm{R}}^d
  u_{\mathrm{R}}\right].
\end{equation}
In terms of $\delta \equiv T_{\mathrm{L}}/T_{\mathrm{R}}-1$, we find
(using results from Appendix \ref{AppendixA}): \begin{equation}
  \mathcal{S} = \frac{\sqrt{d}\left(d^2-1\right)}{48}CT_{\mathrm{R}}^d
  \delta^3 + \mathrm{O}\left(\delta^4\right).
  \label{eq:entprod}
\end{equation}
As in the non-relativistic case \cite{landau}, $\mathcal{S}$ vanishes
at leading order in $\delta$, and only appears at order $\delta^3$.
We can relate $\mathcal{S}$ to the width of the shock by noting that
in a conformal fluid, the only source of entropy production (at
leading order) is through viscous dissipation, so a scaling argument
immediately leads to \begin{equation} {\mathcal S}_{\text pert}\sim
  \int\limits_{\mathrm{shock}} \mathrm{d}x \; \frac{\eta}{T}
  \left(\partial_x v_x\right)^2.
\end{equation}
Evaluating this on the perturbative solution corresponding to a
gaussian profile with width $l_{\text shock}\sim \sqrt{Dt}$ one
obtains
\begin{equation}
{\mathcal S}_{\text pert}\sim \frac{\eta v_{\mathrm{h}}^2}{Tl_{\mathrm{shock}}}.
\end{equation}
At late times this entropy production rate is not sufficient to be
compatible with Eq.~(\ref{eq:entprod}). We conclude that perturbation
theory breaks down at a characteristic timescale
\begin{equation} t_{\mathrm{shock}} \sim
  \frac{\eta}{T_{\mathrm{R}}^{d+1}\delta^2} ,
\end{equation}
making it clear that this effect is non-perturbative in $\delta$. This
effect cannot be seen by directly solving the hydrodynamic equations
perturbatively with the step (Riemann) profile. In non-relativistic
fluids, it is typically the case that the shock simply stops growing,
and maintains a finite width, similar to a soliton \cite{landau}. It
would be interesting to confirm this for the relativistic fluid.

In the regime where $T_{\mathrm{L}} \gg T_{\mathrm{R}}$, the above
argument breaks down.  Noting by dimensional analysis that $\eta \sim
T^d$, we estimate that perturbation theory breaks down
when \begin{equation} l_{\mathrm{shock}} \sim
  \frac{1}{T_{\mathrm{L}}}.
\end{equation}
It would be interesting to study this problem more carefully in future
work, most likely through numerical simulations.  Since our estimate
of $l_{\mathrm{shock}}$ is comparable to the scale at which
hydrodynamics itself breaks down, higher derivative corrections to the
hydrodynamic equations cannot be neglected.  This work could
potentially be carried out using gauge-gravity duality, as this
holographic approach automatically ``resums" hydrodynamics to all
orders in the gradient expansion.

\section{Momentum Relaxation}
\label{Sec:Momentum}

In the previous sections, we have focused on fluids without impurities
or other lattice effects which break translation invariance.  In many
realistic physical systems (such as electron fluids in metals), these
effects are present, but if weak, they may be systematically accounted
for within a hydrodynamic framework.  As these effects violate
momentum conservation,\footnote{The heat current, not energy, is
  exactly conserved \cite{lucas}; however, upon spatial averaging this
  effect is not important.}  we may extend the hydrodynamic equations
(\ref{eq:conslaws}) on very long wavelengths
to: \begin{subequations}\label{eq:momreleqs}\begin{align} \partial_\mu
    T^{\mu t} &= 0, \\ \partial_\mu T^{\mu i} &= -\frac{
      T^{ti}}{\tau}.
\end{align}\end{subequations}
Here, $\tau$ is a phenomenological parameter corresponding to the time
scale over which momentum decays.  The validity of this hydrodynamic
approximation has been shown explicitly in the limit where the fluid
velocity is small compared to the speed of light by coupling the fluid
to sources which break translational symmetry \cite{lucas}.  However,
Eq.~(\ref{eq:momreleqs}) has been used for quite some time (see
e.g. \cite{hkms, davison, blake}).  When the rate is small and
momentum relaxation is weak, the stress tensor of the fluid will be
approximately unchanged from the clean fluid \cite{lucas}, and we may
continue to use the stress tensor (\ref{eq:tmunu}).  The validity of
(\ref{eq:momreleqs}) for flow velocities comparable to the speed of
light is less clear, but as we'll see, the fluid velocities tend to be
small at late times. We therefore expect that our discussion is
qualitatively right.

\begin{figure}[t]
\centering 
\includegraphics[width=3.5in]{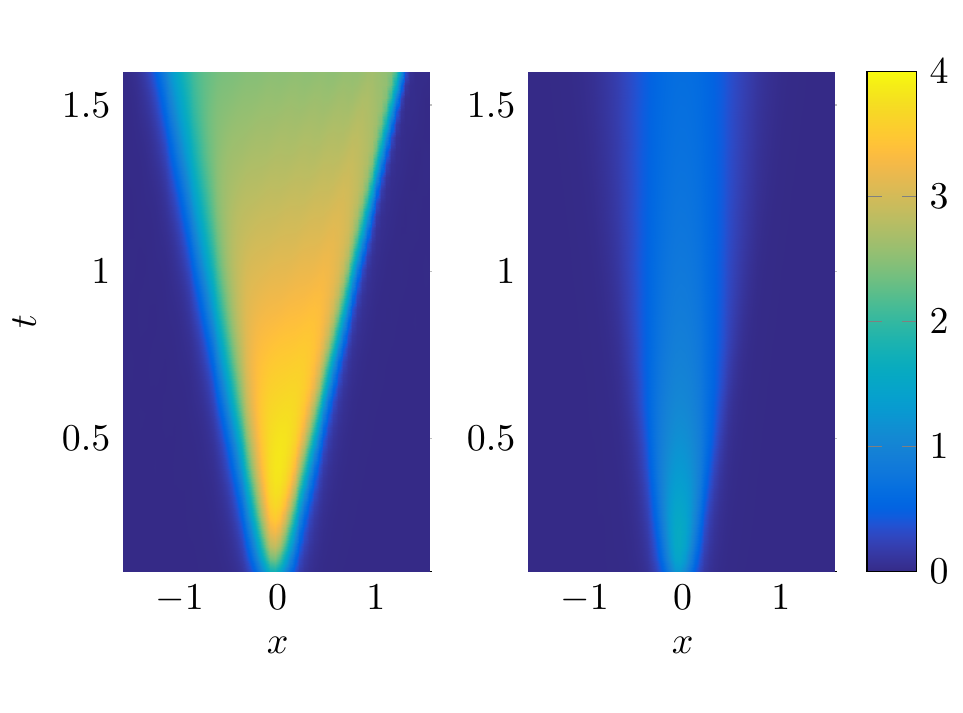}
\caption{An intensity plot of $T^{tx}$ as a function of $x$ and $t$,
  in conformal hydrodynamics with momentum relaxation.  We use the
  same initial conditions as in Fig.~\ref{fignum}, but with
  $T_{\mathrm{L}}=2$.  The left panel shows $\tau=3$, and the right
  panel shows $\tau =0.3$. The value of the energy current, and the
  width of the region between the shock waves, is evidently reduced in
  the right panel. }
\label{fig6}
\end{figure}

\begin{figure}[t]
\centering \includegraphics[width=3.5in]{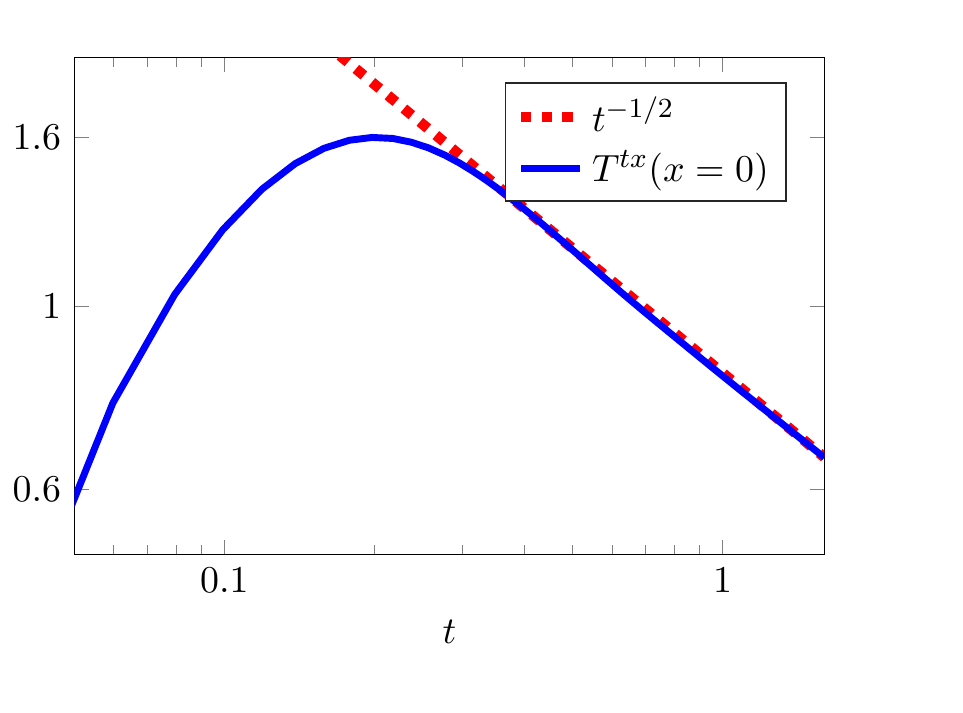}
\caption{The algebraic decay of the energy current at the interface
  obtained by conformal hydrodynamics with momentum relaxation.  We
  use the same initial conditions as in Fig.~\ref{fig6} and set
  $\tau=0.3$.}
\label{fig7}
\end{figure}

As we discussed earlier for the Riemann problem, the temperatures in
the problem do not introduce a relevant time scale for hydrodynamic
phenomena.  At the perfect fluid level, the only time scale in the
problem is $\tau$.  For $t\ll \tau$, the dynamics of the fluid is
effectively described by the solution of Section
\ref{Sec:Rarefaction}.  For $t\gg \tau$, if the system reaches a
steady state where $\partial_t$ and $\partial_x$ are ``small", the
energy flow is determined by the equation
\begin{equation} T^{tx}
  \approx -\tau \partial_x T^{xx}.  \label{eq:mr1}
\end{equation}
Hence, if a steady state forms, the fluid velocity and $T^{tx}$ vanish
as $t\rightarrow \infty$, yielding an equilibrium state with
\begin{equation}
 T^{tt} \approx d T^{xx}.  \label{eq:mr2}
\end{equation}
Hence, the momentum-relaxing hydrodynamic equations lead to a
diffusion equation for the energy density $\epsilon = T^{tt} =
dCT^{d+1}$ (and pressure) for $t\gg \tau$: \begin{equation} \partial_t
  \epsilon \approx \frac{\tau}{d}\partial_x^2 \epsilon,
\end{equation}
where the diffusion constant is $\tau c_{\mathrm{s}}^2$; since this is
perfect conformal hydrodynamics, the speed of sound is
$c_{\mathrm{s}}^2=1/d$.  Remarkably, although the dynamics for $t\ll
\tau$ is highly nonlinear, momentum relaxation reduces the late time
dynamics to simple diffusion.  We conclude that when $t\gg
\tau$: \begin{equation} \epsilon = dC
  \left[\frac{T_{\mathrm{R}}^{d+1}+T_{\mathrm{L}}^{d+1}}{2} +
    \frac{T_{\mathrm{R}}^{d+1}-T_{\mathrm{L}}^{d+1}}{2}
    \mathrm{erf}\frac{x}{\sqrt{t\tau/d}}\right].  \label{eq:mr3}
\end{equation}
This late time behavior can also be seen by considering the fate of
sound modes in the linear response regime \cite{davison}.
Fig.~\ref{fig6} shows a numerical simulation of the equations of
momentum relaxing hydrodynamics.  It is clear that the NESS does not
persist for times $t\gtrsim \tau$.
Indeed, as $t\rightarrow \infty$, $ T^{tx} \sim t^{-1/2}$, as is readily
shown from Eqs~(\ref{eq:mr1}), (\ref{eq:mr2}) and (\ref{eq:mr3}).  We
have confirmed this numerically in Fig.~\ref{fig7}.  Experimental
observation of the NESS thus requires probing the quantum dynamics of
these inhomogeneous systems on fast time scales compared to $\tau$.

\section{Charged Fluids}
\label{Sec:Charge}

A straightforward extension of our results in $d>1$ is to quantum
critical systems with a conserved charge.  In that case, the two
asymptotic heat baths may have different chemical potentials
$\mu_{\mathrm {L,R}}$, in addition to different temperatures
$T_{\mathrm {L,R}}$.  Eq. (\ref{eq:conslaws}) is then supplemented by
charge conservation, \begin{equation} \partial_\mu J^\mu = 0.
\end{equation}
For a perfect fluid, \begin{equation}
J^\mu = nu^\mu,
\end{equation}
where $n$ is the charge density.  In a relativistic gapless fluid,
Eq.~(\ref{eq:tmunu}) is unchanged (see e.g. \cite{kim2}), up to
replacing $CT^{d+1}$ with $P(\mu, T)$, the pressure in the local fluid
rest frame.  Hence, the dynamics of $P$ and $u^\mu$ closes and
decouples from the dynamics of $n$.  The energy-momentum dynamics is
therefore the same as described in Section \ref{Sec:Rarefaction},
after making the replacements $CT_{\mathrm{L,R}}^{d+1} \rightarrow
P_{\mathrm{L,R}}$, where $P_{\mathrm{L,R}}$ denote the pressures in
the left and right baths at $t=0$.

In the rarefaction wave, a straightforward analysis similar to that
for energy and momentum gives us the local charge density
$n(x,t)=n(\xi)$. Clearly to the left of the rarefaction wave it is
identical to the left asymptotic bath value, and to the right of the
shock wave it equals the right asymptotic bath value.  Within the
left-moving rarefaction wave, \begin{equation} n(\xi) = n_{\mathrm{L}}
  \left(\frac{1-v}{1+v}\right)^{\sqrt{d}/2}
\end{equation}
where $n_{\mathrm{L,R}}$ are the initial charge densities in the
left/right reservoirs.  At the right edge of the rarefaction wave, we
have \begin{equation} n=n_{\mathrm{L,h}} \equiv n_{\mathrm{L}}
  \left(\frac{1-v_{\mathrm{h}}}{1+v_{\mathrm{h}}}\right)^{\sqrt{d}/2}.
\end{equation}    
Just to the left of the right shock wave $n = n_{\mathrm{R,h}}$, with
$n_{\mathrm{R,h}}$ given by a Rankine-Hugoniot
equation: \begin{equation}
  \frac{n_{\mathrm{R,h}}v_{\mathrm{h}}}{\sqrt{1-v_{\mathrm{h}}^2}}-u_{\mathrm{R}}
  \frac{n_{\mathrm{R,h}}}{\sqrt{1-v_{\mathrm{h}}^2}}= -u_{\mathrm{R}}
  n_{\mathrm{R}}. \label{eq:nRH}
\end{equation} 

\begin{figure}[t]
\centering 
\includegraphics[width=6in]{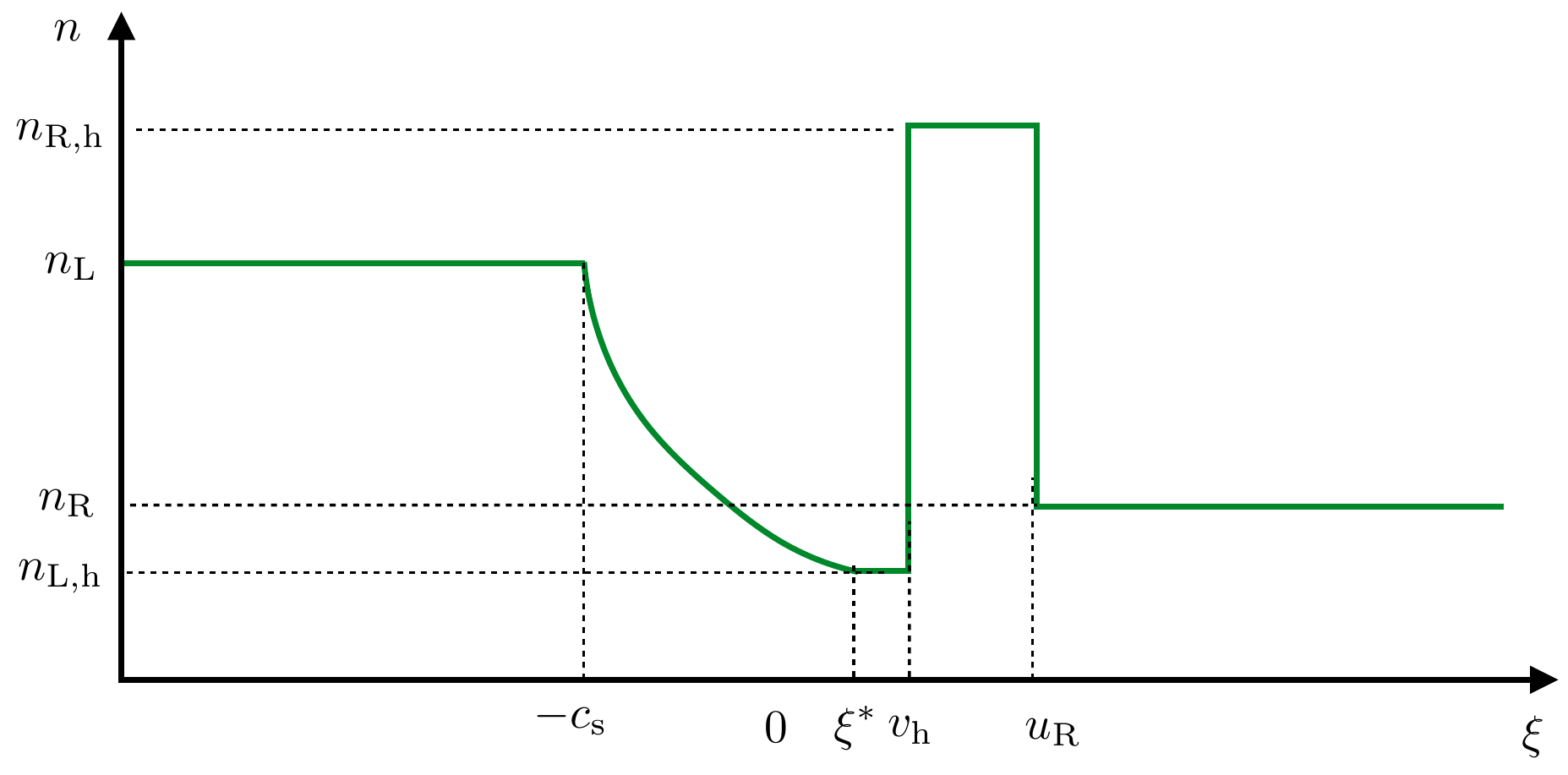}
\caption{A qualitative sketch of the charge density $n(\xi)$ in the
  hydrodynamic Riemann problem for a charged fluid without momentum
  relaxtion.}
\label{figcharge}
\end{figure}

For generic values of $n_{\mathrm{L,R}}$, it will be the case that
$n_{\mathrm{L,h}} \ne n_{\mathrm{R,h}}$. A new shock wave appears
where the charge density jumps between these two values; see
Fig.~\ref{figcharge}.
Such a shock must move at $v_{\mathrm{h}}$, the velocity of the fluid
in the steady state.\footnote{This is commonly called a contact
  discontinuity in the literature on non-relativistic shocks.  The
  presence of quantum critical charge diffusive processes \cite{hkms}
  means that unlike for a Galilean invariant fluid, entropy
  \emph{will} be produced at this discontinuity (charge diffusion
  occurs without fluid flow).  However, this entropy production cannot
  be computed in the ideal fluid limit, as it could for the
  right-moving shock wave, and so the rate of entropy production
  likely vanishes algebraically with $t$.  In the linear response
  regime, this decay rate is $t^{-1/2}$.}  This follows directly from
studying the charge conservation equation in the local rest frame of
the fluid, in the uniform boosted region between the rarefaction and
shock waves.  This slowly moving shock wave will also exhibit
diffusive broadening, analogous to the discussion in Section
\ref{Sec:Viscous}.  Only at late times will this diffusive correction
to the NESS be convected away; hence, numerically detecting this NESS
may require some care.  In the special case where $P_{\mathrm{L}} =
P_{\mathrm{R}}$, the dynamics is entirely governed by (nonlinear)
charge diffusion.  In this case we do not expect a NESS with a
non-zero charge current to appear as $t\rightarrow \infty$; with
$P_L=P_R$ the energy current and the fluid velocity is zero, hence
$J^x=nu^x=0$.

For a discussion of entropy balance for the right-moving shock wave,
see Appendix~\ref{AppendixB}.


\section{Discussion and Conclusions}
\label{Sec:Conc}

In this manuscript we have examined the non-equilibrium energy flow
between quantum critical heat baths in arbitrary dimensions. We have
shown that it is necessary to consider both shock waves and
rarefaction waves in order to describe the steady state energy flow in
$d>1$. This yields minor corrections to the numerical value of
$J_{\mathrm{E}}$ in the resulting NESS, compared to our previous work
\cite{Bhaseen:Energy}. However, there is a qualitative change in the
approach to the NESS for large temperature ratios $T_L/T_R>\Gamma$,
where $\Gamma\approx 3.459$ in $d=2$ and $\Gamma\approx 2.132$ in
$d=3$. We have also discussed extensions of our previous work to
account for viscous broadening of shock waves, and the generalization
of our hydrodynamic solution to inhomogeneous fluids as well as
charged fluids. Although the exact analytical characterization of the
NESS presented in \cite{Bhaseen:Energy} has been modified in $d>1$,
other aspects of the hydrodynamic discussion -- including the
robustness of $J_{\mathrm{E}}$ against perturbations inhomogeneous in
the transverse spatial directions \cite{Bhaseen:Energy} -- are
unchanged.

Though our focus in this paper has been on the appearance of a
non-equilibrium steady-state, we hope to return to the quantum and
thermal fluctuations of this energy current, captured by higher point
correlation functions of $J_{\mathrm{E}}$.  In the idealized two-shock
solution we showed that all higher order moments of the (total) energy
current in the NESS are recursively related to the average (total)
current (across the contact interface). These {\em extended
  fluctuation relations} (EFR) are exact in 1+1 dimensional conformal
theories \cite{Doyon:Heat}, and are asymptotically correct in free
field theories in $d>1$ \cite{Bernard:Time, doyonfree}, where the
dynamics is qualitatively similar to rarefaction waves.  It will be
interesting to revisit the arguments for deriving the EFRs in higher
dimensional systems, presented in [12], in the light of our new
hydrodynamic results for $T_{\rm L}/T_{\rm R}>\Gamma$.

\textbf{Note added:} While this manuscript was in preparation, we were
informed about similar results obtained in \cite{SpillaneHerzog}.

\acknowledgments

\noindent
We thank P. Romatschke for pointing out Ref. \cite{MartiMueller}.  AL
is supported by the NSF under Grant DMR-1360789 and MURI grant
W911NF-14-1-0003 from ARO. KS is supported by a VICI grant of the
Netherlands Organization for Scientific Research (NWO), by the
Netherlands Organization for Scientific Research/Ministry of Science
and Education (NWO/OCW) and by the Foundation for Research into
Fundamental Matter (FOM). MJB thanks the Thomas Young Center and the
EPSRC Centre for Cross-Disciplinary Approaches to Non-Equilibrium
Systems (CANES) funded under grant EP/L015854/1.

\begin{appendix}

\section{Perturbative Comparison of Solutions}\label{AppendixA}

In this Appendix we compute the properties of the NESS to third order
in perturbation theory as a function of the perturbative
parameter \begin{equation} \delta \equiv
  \frac{T_{\mathrm{L}}}{T_{\mathrm{R}}}-1.  \label{deltaeq}
\end{equation}
One finds for the rarefaction-shock solution that 
\begin{subequations}\begin{align}
    T_{\mathrm{h}} &= T_{\mathrm{R}}\left[1+\frac{\delta}{2} - \frac{\delta^2}{8} + \frac{23+2d-d^2}{384}\delta^3\right] + \mathrm{O}\left(\delta^4\right),
 \label{rareshock}\\
v_{\mathrm{h}} &= \sqrt{d}\left[\frac{\delta}{2} - \frac{\delta^2}{4} + \frac{65-18d+d^2}{384}\delta^3\right]  + \mathrm{O}\left(\delta^4\right).
\end{align}\end{subequations}
For the two-shock solution, one finds instead \begin{subequations}\begin{align}
T_{\mathrm{h}}^{\text{2-shock}} &= T_{\mathrm{R}} \left[1 + \frac{\delta}{2} - \frac{\delta^2}{8} + \frac{\delta^3}{16} \right] + \mathrm{O}\left(\delta^4\right), \\
v_{\mathrm{h}}^{\text{2-shock}} &= \sqrt{d} \left[\frac{\delta}{2} - \frac{\delta^2}{4} + \frac{33-10d+d^2}{192} \delta^3\right] + \mathrm{O}\left(\delta^4\right).
\end{align}\end{subequations}
Note that deviations between the two solutions occur only at
$\mathrm{O}(\delta^3)$, and for $d>1$. In addition, the change in the
coefficients is quite small for the physical dimensions of $d=2,3$.

An alternative way to write these equations is to note that
$T_{\mathrm{h}}^{\text{2-shock}}=\sqrt{T_{\mathrm L}T_{\mathrm R}}$
\cite{Bhaseen:Energy}. Using this in (\ref{rareshock}) we may recast
the rarefaction-shock solution in the form
\begin{equation}
T_{\mathrm h}=\sqrt{T_{\mathrm L}T_{\mathrm
    R}}\left(1-\frac{(d-1)^2}{384}\delta^3+\mathrm{O}(\delta^4)\right).
\end{equation}
It is readily seen that $T_{\mathrm h}=\sqrt{T_{\mathrm L}T_{\mathrm
    R}}$ in $d=1$, but it receives cubic corrections in $\delta$ for
$d\neq 1$. Similarly,
\begin{equation}
v_{\mathrm h}=v_{\mathrm h}^{\text{2-shock}}
\left(1-\frac{(d-1)^2}{192}\delta^2+\mathrm{O}(\delta^3)\right).
\end{equation}
Again, the results for the rarefaction-shock and two-shock solutions
coincide in $d=1$, but quadratic corrections in $\delta$ appear for
$d\neq 1$.

\section{Entropy Change Across The Right-Moving Shock}\label{AppendixB}
In this Appendix, we demonstrate that the entropy change across the
right-moving shock remains compatible with the second law of
thermodynamics, even in the presence of charge degrees of
freedom. Specifically, we show that the entropy density just to the left
of the right-moving shock, $s_{\mathrm{R,h}}$, is larger than the
entropy density of the right heat bath, $s_{\mathrm{R}}$, in a scale
invariant relativistic charged fluid; see Figs \ref{fig4} and
\ref{figcharge}. In this pursuit, we first review some thermodynamic
preliminaries.

  In equilibrium, the entropy density is constrained by scale
  invariance and dimensional analysis to have the form
    \begin{equation} s = \epsilon^{d/(d+1)}
  f\left(\frac{n}{\epsilon^{d/(d+1)}}\right),  \label{eq:b1}
  \end{equation}
  where $\epsilon$ is the energy density, $n$ is the charge density
  and $f$ is a function of the dimensionless ratio  \begin{equation}
  \mathcal{X} = n\epsilon^{-d/(d+1)}.
  \end{equation}
  Although the function $f$ is specific to
  the model under consideration, it satisfies some general
  properties. For example, charge conjugation symmetry about $n=0$
  implies that $f(\mathcal{X}) = f(-\mathcal{X})$. Further, using the first law of thermodynamics
  \begin{equation} \mathrm{d}s = \frac{1}{T} \mathrm{d}\epsilon
    - \frac{\mu}{T}\mathrm{d}n, \label{eq:1stlaw}
\end{equation}
we see that $(\partial s/\partial n)_\epsilon = - \mu/T$.  Using
charge conjugation symmetry, we find $f^\prime(0)=0$, as $\mu=0$ when
$n=0$.  More generally, we conclude that $f^\prime(\mathcal{X})>0$ if
$\mathcal{X}<0$ and $f^\prime(\mathcal{X})<0$ if $\mathcal{X}>0$,
corresponding to $f(\mathcal{X})$ having a maximum at
$\mathcal{X}=0$.\footnote{This is consistent with the linear response
  relation $D=-T\sigma_{\textsc{q}}\partial_n^2 s$, where
  $\sigma_{\textsc{q}}$ is a positive dissipative hydrodynamic
  coefficient.  Positivity of the charge diffusion constant $D$ requires
  $\partial_n^2 s < 0$ and thus $f^{\prime\prime}(\mathcal{X})<0$.
  Hence $f(\mathcal{X})$ has a maximum at $\mathcal{X}=0$.  }

\begin{figure}[t]
\centering
\includegraphics[width=3.2in]{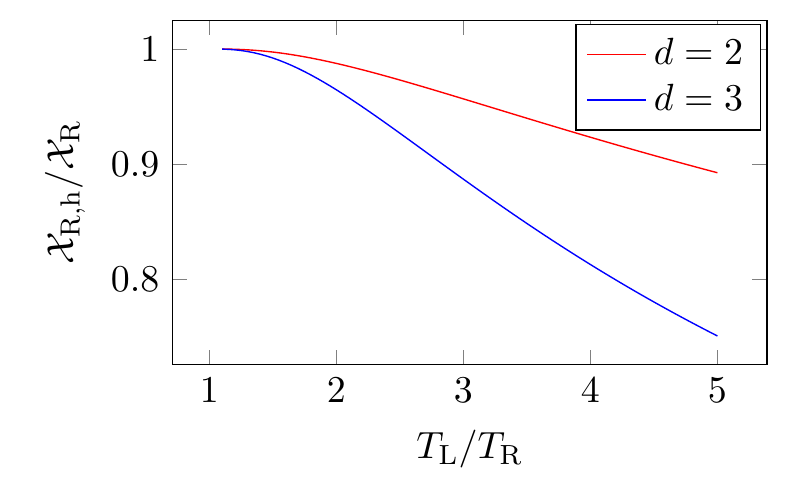}
\caption{A comparison of the dimensionless ratios $\mathcal{X}_{\mathrm{R,h}}$ and
  $\mathcal{X}_{\mathrm{R}}$ to the left and the right of the right-moving
  shock. The ratio $\mathcal{X}_{\mathrm{R,h}}/\mathcal{X}_{\mathrm{R}}$ is always less than
  unity for $d=2,3$.}
\label{fig:lambdas}
\end{figure}

Without loss of generality we may take $\mathcal{X}_{\mathrm{R}}>0$,
due to charge conjugation symmetry.  In order to show that
$s_{\mathrm{R,h}}-s_{\mathrm{R}}\ge 0$, it is sufficient to show that
$f(\mathcal{X}_{\mathrm{R,h}})\ge f(\mathcal{X}_{\mathrm{R}})$,
since $\epsilon_{\mathrm{R,h}}>\epsilon_{\mathrm{R}}$. In
Fig.~\ref{fig:lambdas} we plot the ratio
$\mathcal{X}_{\mathrm{R,h}}/\mathcal{X}_{\mathrm{R}}$ in $d=2,3$, as
obtained from Eqs.~(\ref{tsr}) and (\ref{eq:nRH}) using our numerical
values for $v_{\mathrm{h}}$ and $u_{\mathrm{R}}$.  It is readily seen
that $\mathcal{X}_{\mathrm{R,h}}/\mathcal{X}_{\mathrm{R}} \le 1$. It
follows that $f(\mathcal{X}_{\mathrm{R,h}})\ge
f(\mathcal{X}_{\mathrm{R,h}})$, and thus
$s_{\mathrm{R,h}}-s_{\mathrm{R}}\ge 0$, as required.


\end{appendix}

\bibliography{rarefactionbib}

\end{document}